\documentclass[prb,twocolumn,showpacs,floatfix]{revtex4}
\usepackage{epsfig,subfigure,amssymb,amscd,amsmath,graphicx}
\usepackage{times}

\newcommand{\nin}{\noindent}
\newcommand{\non}{\nonumber}

\newcommand{\bea}{\begin{eqnarray}}
\newcommand{\eea}{\end{eqnarray}}
\newcommand{\be}{\begin{equation}}
\newcommand{\ee}{\end{equation}}

\newcommand{\ket}[1]{     |    \,    #1    \rangle}
\newcommand{\bra}[1]{  \langle #1  \,  |} 
\newcommand{\ZZ}{\mathbb{Z}}

\newcommand{\anb}{b^{\phantom\dagger}}
\newcommand{\crb}{b^\dagger}
\newcommand{\anc}{c^{\phantom\dagger}}
\newcommand{\crc}{c^\dagger}

\newcommand{\bk}{{\boldsymbol{k}}}

\newcommand{\bq}{{\boldsymbol{q}}}

\newcommand{\bs}[1]{ \boldsymbol{#1} }

\begin{document}

\title{A description of Kitaev's honeycomb model with toric code stabilizers}

\author{G. Kells$^{1}$, J. K.~Slingerland$^{1,2}$ and J. Vala$^{1,2}$}


\affiliation{$^{1}$  Department   of  Mathematical  Physics,  National University of Ireland, Maynooth, Ireland, \\ $^{2}$ Dublin Institute for Advanced  Studies, School of Theoretical  Physics, 10 Burlington Rd, Dublin, Ireland. }

\begin{abstract}
We present a solution of Kitaev's spin model on the honeycomb lattice and of related topologically ordered spin models. We employ a Jordan-Wigner type fermionization and find that the Hamiltonian takes a BCS type form, allowing the system to be solved by Bogoliubov transformation. Our fermionization does not employ non-physical auxiliary degrees of freedom and the eigenstates we obtain are completely explicit in terms of the spin variables. The ground-state is obtained as a BCS condensate of fermion pairs over a vacuum state which corresponds to the toric code state with the same vorticity. 
We show in detail how to calculate all eigenstates and eigenvalues of the model on the torus. In particular, we find that the topological degeneracy on the torus descends directly from that of the toric code, which now supplies four vacua for the fermions, one for each choice of periodic vs.~anti-periodic boundary conditions. The reduction of the degeneracy in the non-Abelian phase of the model is seen to be due to the vanishing of one of the corresponding candidate BCS ground-states in that phase. This occurs in particular in the fully periodic vortex-free sector. The true ground-state in this sector is exhibited and shown to be gapped away from the three partially anti-periodic ground-states whenever the non-Abelian phase is gapped. 
\end{abstract}


\date{\today} \maketitle

\section{Introduction}

A combination of special properties has made Kitaev's spin model on the honeycomb lattice~\cite{Kit06} a very popular subject of study in recent years. The model has a basic Hamiltonian with only nearest neighbour interactions, but nevertheless, extensions of the model with magnetic field like terms have both Abelian and non-Abelian topological phases. Moreover these extended models can still be exactly solvable, allowing in principle for direct study of both phases and of the phase transition.   

In spite of the availability of what is by now a large collection of exact solutions \cite{Kit06,Chen07b,Yu08,Pac07,Lah07,Bas07,Feng07,Chen07a,Lee07,Seng08,Yang08, Gu08,Zhao08,Nuss08,Lah09}, all based on fermionization techniques, the two types of phases of the system are generally understood through the use of different methods of analysis. On the one hand Kitaev showed\cite{Kit06}  using perturbation theory that the low energy effective theory of the Abelian phases is equivalent to his $\ZZ_2$ toric code model\cite{Kit03} and in the mean time, extensive further perturbative work on this phase has been done \cite{Sch07,Dus08,Vid08,Kel08a,Kel08b}. On the other hand, the non-Abelian phase is only understood using the fermionized exact solutions of the system. Its topological order is known to be described by the Ising model of topological field theory. 

There have been a number of recent works linking the $\ZZ_2$ toric code  and Ising topological field theories directly.  For example it was demonstrated that a $\ZZ_2$ toric code theory could be formed by condensing bosonic excitations in a doubled Ising theory\cite{Bais08a, Bais08b}. Also, excitations with properties of Ising anyons were constructed from superpositions of the electric and magnetic excitations of the toric code model\cite{Woo08}. In the present work, we want to explore the relationship between these two types of topological order within the context of the honeycomb model.

In order to do this, we introduce yet another solution of the model, but one which is particularly useful for studying the relation between the toric code and Ising type topological orders which exist in the model. Our solution is again by a Jordan-Wigner type fermionization, however, the Jordan-Wigner transformation we employ is closely linked to a choice of basis for the Hilbert space adapted to perturbative analysis of the Abelian, toric code type phase\cite{Sch07,Dus08,Vid08}. The fermions we use are also closely related to the deconfined fermionic excitations which were shown to occur throughout the phase diagram in ref.~\onlinecite{Kel08a} and which correspond to the fermionic excitations of the toric code in the Abelian phase.  After fermionization, the model can be solved exactly and, as with other fermionization methods (cf.~refs.~\onlinecite{Chen07b,Yu08}), the ground-state sector of the system can be transformed to that of a spinless p-wave superconductor, as analyzed by Read and Green in ref.~\onlinecite{Read00}. The ground-state is thus a BCS type state \cite{BCS57}, and can be related to the $\nu =5/2$ fractional quantum Hall state of Moore and Read \cite{Moo91}. With our method, we obtain a vacuum for the fermionized theory which is exactly defined in terms of toric code stabilizers and independent of the couplings of the model. The ground-state for the full system, valid for all parameter space, is in fact a BCS type condensate over the toric code ground-state. Because the vacuum is independent of the coupling parameters, the mechanism for switching between topological phases is contained exclusively within the BCS product. On the other hand, the topological degeneracies of the model are already present at the level of the toric code vacuum. The BCS product only lifts some of this degeneracy in the non-Abelian phase. 

The structure of the paper is as follows:
We start with a short review of the model and of the emergence of the toric code as an effective description of the Abelian phase. We give special attention to a description of the Hilbert space of the model in terms of hard core bosons and effective spins on a square lattice, as this is essential preparation for our fermionization scheme. In section~\ref{sec:plane} we fermionize and solve the model on the plane and give an explicit expression for the ground-state of the model in eq.~(\ref{eq:BCS}). This expression involves only the physical degrees of freedom of the model; no auxiliary variables are introduced anywhere in this work. 
In section~\ref{sec:torus}, we extend our fermionization method to the torus. We explain how to construct the eigenstates in any given vorticity sector and  how to calculate their corresponding energy eigenvalues. Since the creation of an odd number of fermions does not preserve vorticity, certain low energy eigenvalues which might be expected do not occur in every vorticity sector. We give particular attention to the vortex free sector which contains the model's ground-states, and show that the energy of the lowest lying fully periodic state in this sector is lifted in the non-Abelian phase, proving that the ground-state of the non-Abelian phase is three-fold degenerate. In appendix~\ref{app:HFB}, we give a general discussion of Hartree-Fock-Bogoliubov theory with gauge violating fermions, which is used as background for the discussion in section~\ref{sec:torus}

\section{Honeycomb model and toric code}

\subsection{Spin Hamiltonian and loop symmetries}
\begin{figure}
\includegraphics[width=.30\textwidth,height=0.2\textwidth]{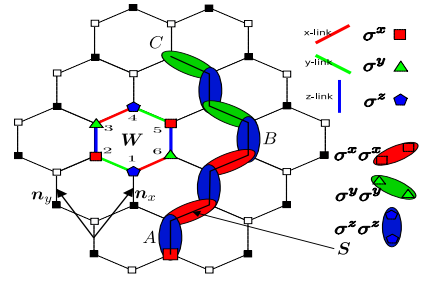}
       \caption{The plaquette operator $\bs{W}$ and the fermionic string
$\bs{S}$}
       \label{fig:hexlattice2}
\end{figure}

The system consists of spins on the sites of a hexagonal lattice. The Hamiltonian can be written as 
\be 
H_0 = - \sum_{\alpha \in \{ x,y,z \}} \sum_{i,j} J_\alpha K_{i,j}^{\alpha} 
\label{eq:H}
\ee 
where $K_{ij}^\alpha = \sigma_i^\alpha \sigma_j^\alpha$ denotes a directional spin exchange interaction occurring between the sites ${i,j}$ connected by a $\alpha$-link see FIG. \ref{fig:hexlattice2}.  We define a the basic unit cell of the lattice with the two unit vectors $\bs{n}_x$ and $\bs{n}_y$ as shown in FIG. \ref{fig:hexlattice2}. By contracting each $z$-link to a single point we define the position vector labeling the z-dimers on a square lattice as $\bs{q}=q_x\bs{n}_x +q_y \bs{n}_y$. 

Consider now products of $K$ operators along loops on the lattice , $K^{\alpha^{(1)}}_{ij} K^{\alpha^{(2)}}_{jk} .... K^{\alpha^{(n)}}_{li}$ , where $\alpha^{(m)} \in {x,y,z}$. Any loop constructed in this way commutes with the Hamiltonian and with all other loops. The shortest such loop symmetries are the plaquette operators
\be
\bs{W}_{\bq} = \sigma_1^z \sigma_2^x \sigma_3^y  \sigma_4^z \sigma_5^x
\sigma_6^y,
\label{eq:Wp}
\ee
where the numbers $1$ through $6$ label lattice sites on single hexagonal plaquette, see FIG. \ref{fig:hexlattice2}. We will use the convention that $\bq$ denotes the $z$-dimer directly below the plaquette. The fact that the Hamiltonian commutes with all plaquette operators implies that we may choose energy eigenvectors $\ket{n}$ such that $W_{\bq}=\bra{n} \bs{W}_\bq \ket{n}=\pm 1$.  If $W_{\bq} = -1$ then we say that the state $\ket{n}$ carries a vortex at $\bq$. When we refer to a particular vortex-sector we mean the subspace of the system with a particular configuration of vortices.  The vortex-free sector for example is the subspace spanned by all eigenvectors such that $W_{\bq} = 1$ for all $\bq$.

On a torus, the plaquette operators are not independent, as they obey $\prod W_{\bq} =I$ where the product is over all $\bq$. There are also two independent homologically non-trivial loop symmetries. To represent these we are free to choose any two closed loop operators that traverse the torus as long as they cannot be deformed into each other by plaquette multiplication. All other homologically non-trivial loop symmetries can be constructed from the products of these two operators and the $N/2-1$ independent plaquette operators, (cf. ref. \onlinecite{Kel08a}). Note that when the torus is specified by periodic boundary vectors $(\bs{x}, \bs{y})$ which are integer multiples of the unit vectors   i.e. $\bs{x} = N_x \bs{n}_x$ and $\bs{y} = N_y \bs{n}_y$, it is natural to use  overlapping products of alternating $z$- and $x$-links ($L^{(x)}_{q_y}=\prod K_{ij}^z K_{jk}^x$)  and alternating $z$- and $y$-links ($L^{(y)}_{q_x}=\prod K_{ij}^z K_{jk}^y$), as homologically non-trivial symmetries. We will generally use the operators $L^{(x)}_{0}$ and $L^{(y)}_{0}$ that run through the origin as the two independent symmetries.

The model contains 4 distinct phases \cite{Kit06}. There are three gapped phases (1) $A_x$ with $J_x>J_y+J_z$,  (2) $A_y$ with $J_y>J_x+J_z$ and (3) $A_z$ with $J_z>J_x+J_y$ with (4) a gapless B phase existing in the parameter space between the three $A$ phases. As each A phase is related to the others by a lattice rotation we confine our analysis to the $A_z$  and $B$ phases with out loss of generality. 

The Hamiltonian (\ref{eq:H}) is often extended to include perturbing terms $H_1$ that 
(i) are sums of $K$ operator products (ii) open a gap in the B-phase (iii) break time-reversal symmetry (T-symmetry), see \cite{Kit06,Pac07,Lah07,Lee07,Yu08} and the general analysis of the link or bond algebras in \cite{Nuss08}. The breaking of T-symmetry is essential for relating the model to chiral p-wave superconductors.  As the procedure we will outline here gives the same physical results as the quoted references for generalised T-symmetry breaking we restrict the explicit calculations to the three-body term studied in references \onlinecite{Kit06,Pac07,Lah07}:
\be 
H_1= - \kappa \sum_{ \bq } \sum_{l=1}^6 P(\bq)^{(l)} 
\ee
with the second summation running over the six terms 
\bea
&& \sum_{l=1}^6 P(\bq)^{(l)} =  \sigma^x_1 \sigma^y_6 \sigma^z_5 +  \sigma^z_2 \sigma^y_3 \sigma^x_4  +\\ \non  && \sigma^y_1 \sigma^x_2 \sigma^z_3 + \sigma^y_4 \sigma^x_5 \sigma^z_6 +   \sigma^x_3 \sigma^z_4 \sigma^y_5 + \sigma^y_2 \sigma^z_1 \sigma^x_6.
\eea

Recently a third type of phase has been discovered in extended honeycomb models, featuring gapped Dirac fermions \cite{Nash09}. The phase is opened by allowing the $J_\alpha$ couplings to vary periodically on the lattice. Although we do not examine this phase in this paper, the methodology employed below can in principle be used.

\subsection{The Toric Code as an effective system}

The Hamiltonian (\ref{eq:H}) can be written in terms of hard-core bosons and effective spins of the z-dimers using the mapping \cite{Sch07}:
\bea 
\label{eq:map}
\ket{  \uparrow_{_\blacksquare} \uparrow_{_\square}}  &=& \ket{\Uparrow,0},
\quad
\ket{\downarrow_{_\blacksquare} \downarrow_{_\square}   } = \ket{\Downarrow,0},
\\
\ket{ \uparrow_{_\blacksquare}  \downarrow_{_\square}  } &=& \ket{\Uparrow,1},
\quad
\ket{  \downarrow_{_\blacksquare} \uparrow_{_\square} } = \ket{\Downarrow,1}.
\non
\eea
The labels on the left hand side indicate the states of the z-dimer in the $S_z$ basis. The first quantum number of the kets on the right hand side represents the effective spin of the square lattice and the second is the bosonic occupation number. The presence of a boson indicates an anti-ferromagnetic configuration of the spins connected by a $z$-link.  

In the $A_z$-phase, the dominance of the $J_z$ means that spins on a z-dimer tend to align in the same direction, and therefore in this limit the presence of bosons is energetically suppressed. A perturbative analysis for the low energy effective Hamiltonian in this regime shows that the first non-constant term, occurring at the 4th order, is 
\be
H_{TC} = - J_{\text{eff}} \sum_\bq \bs{Q}_q \otimes I
\ee 
with $\bs{Q}_{\bq}= \tau_{\bs{q}}^z   ~~\tau_{\bs{q}+\bs{n}_x}^y  \tau_{\bs{q}+\bs{n}_y}^y \tau_{\bs{q}+\bs{n}}^z$ where $\tau_\bq^a$ is the Pauli operator acting on the effective spin at position $\bq$ and  $J_{\text{eff}} = \frac{J_x^2 J_y^2 }{ 16 |J_z|^3}$ \cite{Kit06} . This effective Hamiltonian, defined now on a square lattice, is unitarily equivalent to what is known as the toric code (TC) \cite{Kit03}. 
 
The operators $\bs{Q}_\bq$, like the plaquette operators $\bs{W}_\bq$, all commute with each other. The eigenvalues of each operator can therefore be used as quantum numbers to specify eigenstates of the system. We write $\ket{\{Q_{\bq}\}}$ where $\{Q_\bq \}$ is a full list of $Q_\bq$ eigenvalues. Excitations of the TC system are made/moved by applying $\tau^z$ and/or $\tau^y$ operators to a site.  In the lattice orientation we use, $\tau^z_\bq$ changes the eigenvalues  $Q_{\bq-\bs{n}_x}$ and  $Q_{\bq-\bs{n}_y}$  while $\tau^y_\bq$ changes the eigenvalues  $Q_{\bq}$ and  $Q_{\bq-\bs{n}_x-\bs{n}_y}$. On even-even lattices, which can be bi-colored, the quasi-particle excitations occur in two types usually labeled $e$ and $m$.  Excitations of the same type are mutually bosonic but excitations of different types display Abelian anyonic statistics. Importantly, pairs of $e$ and $m$ particles behave as fermions.

In the language of the stabilizer formalism , see \cite{Gott97,Niel00},  we say that the TC states are stabilized by the operators $\bs{Q}_{\bq'} \ket{\{Q_{\bq}\}} =  Q_{\bq'} \ket{\{Q_{\bq}\}}$.  On a plane for example, the TC ground-state is the state such that $\bs{Q}_{\bq'} \ket{\{Q_{\bq}\}} = \ket{\{Q_{\bq}\}}$ for all $\bq$. However, despite their simple description in terms of the stabilizer formalism, it is important to recognize that the TC states are structurally non-trivial and display unusual entanglement and geometric properties. It has been shown for example that the ground-state of the TC system is a Projected Entangled Pair State (PEPS) with virtual dimension D=2 (see for example ref. \onlinecite{Agu07}),  and can also be described in terms of string-net condensates and loop models \cite{Lev05,Fend08}. 


\subsection{Hard-core bosons and stabilizers}

The basis (\ref{eq:map}) also describes anti-ferromagnetic configurations of the z-dimers through the bosonic occupation number and forms an orthonormal basis for the full honeycomb system.  The Pauli operators of the original spin Hamiltonian can be written as (see references \onlinecite{Sch07,Dus08,Vid08}) :
\begin{equation}
    \label{eq:mapping}
    \begin{array}{lcl}
    \sigma_{\bq,_\blacksquare}^x=\tau_\bq^x  (\crb_\bq+ \anb_\bq) &,& 
\sigma_{\bq,_\square}^x=\crb_\bq+\anb_\bq  ,\\  %
    \sigma_{\bq,_\blacksquare}^y=\tau_\bq^y (\crb_\bq+\anb_\bq) &,& 
\sigma_{\bq,_\square}^y= i \,  \tau^z_\bq (\crb_\bq-\anb_\bq), \\ 
\sigma_{\bq,_\blacksquare}^z=
    \tau_\bq^z  &,&  \sigma_{\bq,_\square}^z=\tau_\bq^z (I-2  \crb_\bq
    \anb_\bq),
    \end{array}
\end{equation}
where $\crb$ and $\anb$ are the creation and annihilation operators for the hard-core bosons. In this representation the Hamiltonian (\ref{eq:H}) becomes 
\bea 
H_0  &=& - J_x \sum_{\bq}  (\crb_{\bq}+\anb_{\bq}) \tau^x_{\bq+\bs{n}_x }
(\crb_{\bq+\bs{n}_x}+\anb_{\bq+\bs{n}_x} )  \non \\ 
&-&  J_y \sum_{\bq} i \tau^z_\bq (\crb_{\bq}-\anb_{\bq}   )
\tau^y_{\bq+\bs{n}_y} ( \crb_{\bq  +\bs{n}_y}+\anb_{\bq+\bs{n}_y} )\non \\ 
&-& J_z \sum_{\bq} (I-2 \crb_{\bq} \anb_{\bq}).
\label{eq:Heb}
\eea
with the perturbative term $H_1$ given pictorially in FIG. \ref{fig:Kitaev_extended}.  The Hamiltonian (\ref{eq:Heb}) has been used  in the gapped $A_z$ phase to perturbatively calculate effective Hamiltonians and other measures, to the 10th order in some cases \cite{Sch07,Dus08,Vid08}.  In the next section we will show how to fermionize this Hamiltonian by attaching string operators to the hard-core bosons. The procedure is much like other Jordan-Wigner type approaches but the operator strings that we choose, will be tailored for this system.

In this representation the plaquette operators (\ref{eq:Wp}) become
\be
\bs{W}_{\bs{q}}=(I-2 \bs{N}_{\bs{q}})(I-2\bs{N}_{\bs{q}+\bs{n}_y})
\bs{Q}_{\bs{q}}
\label{eq:WQ}
\ee
where  $\bs{N}_{\bs{q}}=  \crb_{\bs{q}}  \anb_{\bs{q}}$. This relation (\ref{eq:WQ}) is very useful because it allows one to write down an orthonormal basis for the full honeycomb system \cite{Vid08}. The basis can be written as $\ket{ \{W_{\bq}\}, \{\bq\}}$, where $\{\bq\}$ lists the sites with non-zero bosonic occupation and the eigenvalues $\{ W_\bq \}$ determine the vortex sector. Note that to determine the structure of the state one still uses the effective operators $\bs{Q}_\bq$ as the stabilizers with the eigenvalues reflecting the vorticity $\{W_{\bq}\}$ through (\ref{eq:WQ}). In the special case where there are no broken dimers we have  $\ket{ \{W_{\bq}\}, \{\emptyset \}} \equiv \ket{ \{Q_{\bq}\}} $ with $W_\bq = Q_\bq$ for all $\bq$. 

To specify a state on a torus we also give two additional quantum numbers associated with the homologically non-trivial loop symmetries. We can normally choose these to be the eigenvalues $l^{(x)}_0$ and $l^{(y)}_0$ of the independent operators $L^{(x)}_0$ and $L^{(y)}_0$ described above and the generic state in this case can be written as $\ket{ \{W_{\bq}\}, l^{(x)}_0, l^{(y)}_0 , \{\bq\}}$ where the list $\{W_{\bq}\} $ contains $N_x N_y -1$ independent $W_{\bq}$'s. 
  
In the next section we will have cause to use a generalisation of the expression (\ref{eq:WQ}) for products of plaquette operators. Of particular importance, because of the conventions used, will be the products arranged vertically on the effective lattice. We have in this case 
\bea 
\label{eq:X}
\bs{X}_{q_x,q_y} &\equiv& \prod_{q_y'=0}^{q_{y-1}} \bs{W}_{q_x,q_y'}  \\ 
&=& \non (I-2 \bs{N}_{q_x,0})(I-2 \bs{N}_{q_x,q_y}) \prod_{q_y'=0}^{q_{y-1}} 
\bs{Q}_{q_x,q_y'} 
\eea
and we see that only the bosonic occupation numbers at the upper and lower left corners of the plaquette product need to be taken into account.

\section{Fermionization}
\label{sec:plane}

We now show how to turn the hardcore bosons into fermions using a new Jordan-Wigner type transformation that is designed for this model and in particular for the basis  $\ket{ \{W_{\bq}\}, \{\bq\}}$ described above. The procedure has a number of advantages over other fermionization techniques. For example, the method does not introduce additional un-physical degrees of freedom like the Majorana approach originally used to solve the problem \cite{Kit06}  but still allows a transparent encoding of the vorticity within the fermionic Hamiltonian. In addition the procedure also reveals much more about the actual eigenstates of the system than previous fermionization methods. We will see for example, like Chen and Nussinov \cite{Chen07b}, that the ground-state of the system is a BCS type product acting on the vacuum. However, our vacuum will be a exactly defined in terms of Toric Code stabilizers meaning that the eigenstates of the system can be written in simple closed form expressions that do not require implicit spectral projection. 

We begin by defining a particular string operator using overlapping products of the $K^\alpha_{ij}$ terms of the original Hamiltonian. The string will serve two purposes: (i) it will break/fix $z$-dimers at a single location $\bs{q}$ of the lattice thereby creating/annihilating hard-core bosons (ii) it will enforce inter-site fermionic commutation relations effectively turning our hard-core bosons into fermions. 

\begin{figure}  
       \includegraphics[width=.40\textwidth,height=0.25\textwidth]{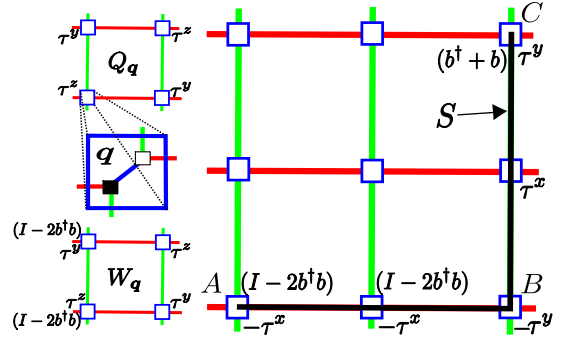}
       \caption{Bosonic and effective spin decomposition of the operator string
S. }
       \label{fig:ExplainS}
\end{figure}

Our convention will be to first apply a single $\sigma^x$ term to a black site of the z-link which we set to be the origin.  The rest of the string is made by applying first alternating $K^z_{ij}$ and $K^x_{jk}$ until we reach a required length and then apply alternating $K^z_{lm}$ and $ K^y_{mn} $ terms ending on the black site at $\bq$,  see FIG. \ref{fig:hexlattice2}. Explicitly we write 
\bea
\label{eq:Sq1}
 S_{\bq} &\equiv&     \sigma^y_{(q_x,q_y),_\blacksquare} 
\sigma^y_{(q_x,q_y-1),_\square} \sigma^z_{(q_x,q_y-1)_\square}  \\
 \non &&...
\sigma^y_{(q_x,1),_\blacksquare} \sigma^y_{(q_x,0),_\square}
\sigma^z_{(q_x,0),_\square}
\sigma^z_{(q_x,0)_\blacksquare} \sigma^x_{(q_x,0)_{\blacksquare}}  \\ \non &&
...   \sigma^x_{(1,0)_\blacksquare}
\sigma^x_{(0,0)_{\square}}\sigma^z_{(0,0),_\square}
\sigma^z_{(0,0),_\blacksquare} \sigma^x_{(0,0),_\blacksquare}.
\eea
Using the representation of \cite{Sch07,Dus08,Vid08} we can decompose (\ref{eq:Sq1}) into the effective spin and bosonic subspaces, i.e. $S=S_{e} \otimes S_b$. In this decomposition there are four different types of structures to observe on the effective lattice: (1) the line including the starting point $A$ up to, but not including, the turning point $B$, (2) the turning point $B=(q_x,0)$,  (3) the exclusive interval $BC$, and (4) the end point $C=(q_x,q_y)$, see FIG \ref{fig:ExplainS} and TABLE \ref{table:Toric-table}.
\begin{table}[ht]
\centering                          
\begin{tabular}{c c l}            
\hline                       
  & S & $~~$ $S_{e} \otimes S_{b}$      \\ [0.5ex]  
\hline                              
[A,B)     & $\sigma^x_{_\square} \sigma^z_{_\square} \sigma^z_{_\blacksquare}
\sigma^x_{_\blacksquare}$ &  $ - \tau^x \otimes I-2 \crb \anb$ 
\\

B    & $\sigma^y_{_\square} \sigma^z_{_\square} \sigma^z_{_\blacksquare}
\sigma^x_{_\blacksquare} $ & $ -\tau^y \otimes I$   \\

(B,C)   & $\sigma^y_{_\square} \sigma^z_{_\square} \sigma^z_{_\blacksquare}
\sigma^y_{_\blacksquare}$ & $~~$ $ \tau^x \otimes I $ \\

C    & $\sigma^y_{_\blacksquare} $ & $~~$ $ \tau^y \otimes \crb +
\anb$ \\[1ex]
\hline         
\end{tabular}         
\caption{The string $S$ as four unique segments.  While bosons are only
created/destroyed at the endpoint $C$ of the string, the  sites in the $[A,B)$
interval also have non-trivial bosonic dependence.
\label{table:Toric-table}  }  
\end{table}

The operator $S_\bq$ squares to unity while different operators $S_\bq, S_{\bq'}$ anti-commute with each other. This leads us to identify the string $S_\bq$ with the following sum of fermionic creation and annihilation operators: $S_{\bq} = c^\dagger_\bq + c_\bq = (\crb_\bq + \anb_\bq)  S_\bq^{'}$ where $S^{'}_\bq$ is simply the string $S_\bq$ but with the bosonic dependence of the end-point $C$ removed,  see TABLE \ref{table:Toric-table}. Individually our fermionic canonical creation and annihilation operators are 
\be
\label{eq:cdef}
\crc_\bq = \crb_\bq S^{'}_\bq, \quad c_\bq = b_\bq S^{'}_\bq
\ee
where the strings now insure that the operators $c_\bq^\dagger$ and $c_\bq$ obey the canonical fermionic anti-commutator relations
\be
\{ c^\dagger_\bq, c^{\phantom \dagger}_{\bq'} \} = \delta_{\bq \bq'}, \;\;\; \{ c^{ \dagger}_\bq ,c^{\dagger}_{\bq'} \}  =0 ,\;\;\; \{ c^{\phantom \dagger}_\bq ,c^{\phantom \dagger}_{\bq'} \} =0 .
\ee

The operators $c^\dagger_\bq$ and $c_\bq$ must both create/annihilate vortices at $-\bs{n}_x$ and $-\bs{n}_x-\bs{n}_y$. We can therefore think of each fermion as being bound to a vortex pair at the origin. Remarkably, this vortex-pair can itself be thought of as a fermion and can even be moved without changing the energy of the system \cite{Kel08a}.  Another interesting insight can be obtained by noting that the strings $S'_\bq$, which we attach to the hard-core bosons to make them fermionic, change the eigenvalues of the $\bs{Q}_\bq$ operators at $-\bs{n}_x$, $-\bs{n}_x-\bs{n}_y$, $\bq$ and $\bq-\bs{n}_y$. The creation of the $c^\dagger$ fermions is thus reflected in the effective spins by the creation of two bound fermionic $e-m$ pairs. These associated states are structurally equivalent, up to the mapping (\ref{eq:map}), to the corresponding TC states.

\begin{figure*}  
       \includegraphics[width=.85\textwidth,height=0.37\textwidth]{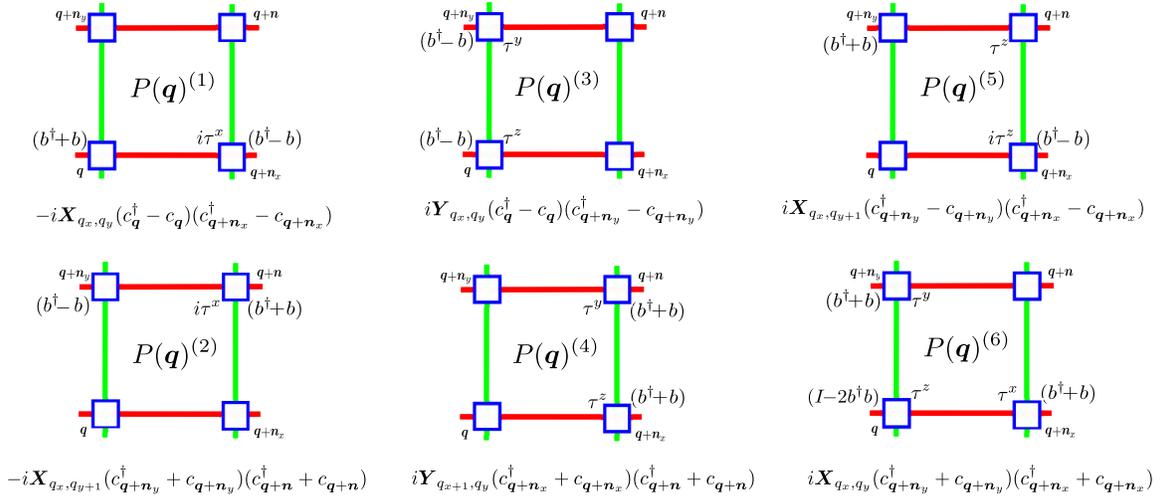}
       \caption{The Kitaev three body term in the effective-spin/hard-core boson and fermionic notation. On a torus the values of $X$ and $Y$ are as those for the basic two-body Hamiltonian described below, except for the terms $P(\bq)^{(5)}$ and $P(\bq)^{(6)}$ when $q_y = N_y$. }
       \label{fig:Kitaev_extended}
\end{figure*}

If we invert (\ref{eq:cdef}) and substitute the relevant expressions into the Hamiltonian (\ref{eq:Heb}) we get
\bea 
\bs{H}_0 &=& J_x \sum_{\bq} \bs{X}_{\bq} (\crc_{\bq}-\anc_{\bq})
(\crc_{\bq+\bs{n}_x}+\anc_{\bq+\bs{n}_x}) \non \\ 
 &+& J_y \sum_{\bq} \bs{Y}_\bq (\crc_{\bq}-\anc_{\bq} ) (
\crc_{\bq+\bs{n}_y}+\anc_{\bq+\bs{n}_y} )\non \\ 
 &+& J_z \sum_{\bq} (2\crc_{\bq} \anc_{\bq} - I) ,
\label{eq:Hc}
\eea
where, in the plane, $Y_\bq= I $ for all $\bq$ and $X_\bq$ is defined in (\ref{eq:X}). The fermionic representation of the perturbative term $H_1$ is given in FIG. \ref{fig:Kitaev_extended}.  We restrict the Hilbert space to the relevant vortex-configuration by replacing $\bs{X}_\bq$ by the eigenvalues $X_\bq$ of that configuration. In the simplest case of the vortex free sector we have $X_\bq =1$ for all $\bq$. This sector, because of a theorem by Lieb \cite{Lieb94}, is known to contain the system ground-state and can be solved exactly in the thermodynamic limit by moving to the momentum representation with the Fourier transform 
\be
c_\bq = M^{-1/2} \sum c_\bk e^{i \bk \cdot \bq}. 
\ee
After substitution into (\ref{eq:Hc}) and anti-symmetrization we have 
\be
H = \sum_\bk \left[ \xi_\bk  c^\dagger_\bk c_\bk +
\frac{1}{2} (\Delta c^\dagger_\bk c^\dagger_{-\bk} + \Delta^* c_{-\bk} c_{\bk}) \right] - MJ_z
\label{eq:Hk}
\ee
where 
\bea
\xi_\bk &=& \varepsilon_\bk -\mu \\
\Delta_\bk &=& \alpha_\bk+i \beta_\bk 
\eea
with
\bea
\mu_{\phantom \bk} &=& -2 J_z \\
\varepsilon_\bk &=&   2J_x \cos(k_x) + 2J_y  \cos(k_y)\\
\alpha_\bk &=&  4 \kappa ( \sin(k_x) -\sin(k_y) -\sin(k_x-k_y)) \\
\beta_\bk &=&  2  J_x \sin(k_x) +  2 J_y\sin(k_y).
\eea
where the extended Hamiltonian $H_1$ is now fully contained within in the $\alpha_\bk$ term. The procedure also gives agreement with the other fermionization techniques to analyse the extended model \cite{Pac07,Lah07,Feng07,Chen07b,Lee07}. We note in particular that the technique can be used to replicate the dispersion relations of \cite{Yu08} where the p-wave pairing can be tuned to have $k_x + ik_y$ chiral symmetry thus allowing a direct link with the work of Read and Green \cite{Read00} and subsequent analysis \cite{Ivanov01,Stern04,Stone04,Stone06}, relating the Pfaffian Quantum Hall states, p-wave superconductors and the Ising CFT model.

The Hamiltonian (\ref{eq:Hk}) is diagonalized by Bogoliubov transformation 
\be
\gamma_\bk = u_\bk \anc_\bk - v_\bk \crc_{-\bk},
\ee
where $u_\bk$ and $v_\bk$ satisfy $|u_\bk|^2+|v_\bk|^2=1$. 
We then have $H=\sum E_\bk (\gamma^{\dagger}_{\bk} \gamma_\bk -1/2)$, with 
\bea
\label{eq:Ek}
E_\bk &=& \sqrt{\xi_\bk^2  +|\Delta_\bk|^2} \\
u_\bk &=& \sqrt{1/2 (1 + \xi_\bk/E_\bk)} \\
v_\bk &=& i\sqrt{1/2(1 - \xi_\bk/ E_\bk)} 
\eea
The ground-state, annihilated by all $\gamma_\bk$, and of energy $E_{\text{gs}}=-\frac{1}{2}\int E_\bk d \bk $,  can be seen to be the BCS type state
\be 
\ket{\text{gs}} = \prod_\bk (u_\bk + v_\bk \crc_\bk \crc_{-\bk} )\ket{\{W_\bq\},\{\emptyset\} }. 
\label{eq:BCS}
\ee

This expression is, to the best of our knowledge, the first closed form expression of the ground-state that does not require additional spectral projection.  It is noteworthy because it combines two powerful wavefunction descriptors i.e. the BCS product and the Stabilizer formalism.  In the expression, which is valid everywhere in the model's parameter space, the fermionic vacuum is fixed to be the toric code ground-state. While this implies that any mechanism for switching between the Abelian and non-Abelian topological phases must be contained exclusively within the BCS product, we should also recognize that the Abelian phase is $\ZZ_2 \times \ZZ_2$ because the fermionic vacuum is $\ZZ_2 \times \ZZ_2$ and not because of any mysterious property of the BCS product.   To see this more clearly note that in the $A_z$ phase with $(J_z=1, J_x,J_y  \rightarrow 0)$ we have $u_\bk \rightarrow 1$ and $ v_\bk \rightarrow 0$ and the the ground-state of the full system  $\ket{\text{gs}} \rightarrow \ket{\{W_\bq\}, \{\emptyset\}} \equiv \ket{\{Q_\bq\}}$, where $Q_\bq =1$ and $W_\bq =1$ for all $\bq$. This is of course what one expects from the perturbation theory, see for example \cite{Kel08a,Kel08b}.

\section{Fermionization on a torus}
\label{sec:torus}

The fermionization procedure above may be extended to systems that live on a torus. 
Going to the torus allows for the study of finite size systems without fixing boundary conditions. It also allows us to probe the topological order of the model's A and B-phases directly. From the predictions of topological-QFT the ground-state degeneracy on a torus should be equal to the number of topological sectors, or quasi-particle types in the system. In the Abelian A-phase we should have a 4-fold degeneracy because we have 4 distinct particle types : the trivial particle or vacuum, the $e$ particle, the $m$ particle, and the fermionic $e$-$m$ composite, which as we have mentioned, corresponds to our $\gamma^{\dagger}$ excitation. In the non-Abelian phase the theory predicts we have three distinct particle types with the distinction between $e$ and $m$ particle types no longer applicable \cite{Kit06}. However, as far as we know, a direct analysis of the ground-state degeneracy in this phase has not been done. This may be because on the surface, it appears that the Read and Green's analysis  for p-wave superconductors in ref.~\onlinecite{Read00} can be carried over directly to this model. However, we will show that this is not the case and that there are a number of subtle differences, the primary one being that our fermions do not preserve the gauge symmetries, i.e.~the creation of a fermion changes the vorticity.  

\begin{figure*}
      \includegraphics[width=.9 \textwidth,height=0.18\textwidth]{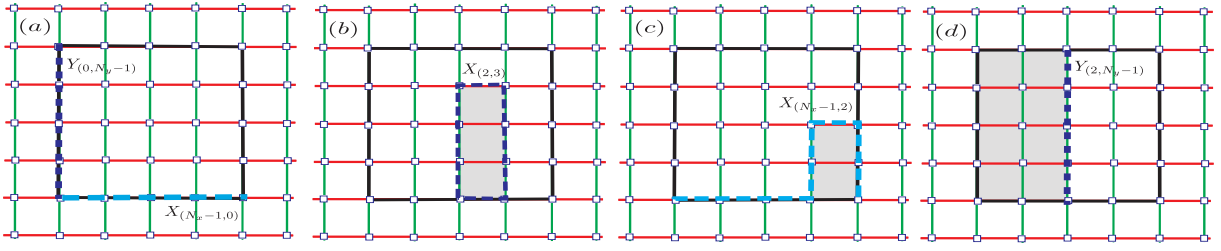}
      \caption{On a torus there are $N/2+1$ independent loop
symmetries \cite{Kel08a}. From these independent loop symmetries all
other loop symmetries, of which there are $2^{N/2+1}$, can be
generated by multiplication. We can specify a particular sector of the
Hamiltonian by specifying the eigenvalues of the $N/2-1$ plaquette
symmetries and two homologically non-trivial loops of our choosing.
The eigenvalues $X_\bq$ and $Y_\bq$ used to specify a particular
sector in (\ref{eq:Hg}) are now fully determined. In the graphs above
we have indicated how a few specific values of $X_\bq$ and $Y_\bq$
depend on the eigenvalues of the independent loop symmetries: (a)
$Y_{(0,N_{y-1})}=-l_0^{(y)}$, $X_{(N_{x-1},0)}=-l_0^{(x)}$ (b) $
X_{(2,3)} = \prod_{q_y=0}^2 W_{(2,q_y)}$ (c) $ X_{(N_{x-1},2)} = -
l^{(x)}_0 \prod_{q_y=0}^1 W_{(N_{x-1},q_y)}$ (d) $Y_{(2,N_{y-1})}=
-l^{(y)}_0 \prod_{q_x =0}^1 \prod_{q_y=0}^{N_y-1} W_{(q_x,q_y)}$ }
      \label{fig:ExplainF}
\end{figure*}

To proceed we first re-write the fermonic Hamiltonian as 
\bea
H=  \frac{1}{2} \sum_{\bq \bq'} \left[\begin{array}{cc} c^\dagger_{\bq} & c_\bq
\end{array}  \right] \left[
\begin{array}{cc} \xi_{\bq \bq'} & \Delta_{\bq \bq'} \\ \Delta^\dagger_{\bq
\bq'} & -\xi^{T}_{\bq \bq'} \end{array} \right] \left[\begin{array}{c}
c_{\bq'}
\\ c^\dagger_{\bq'} \end{array}  \right] 
\label{eq:Hg} 
\eea
where for example with $H_0$ we  would have
\bea 
\non \xi_{\bq \bq'} &=& 2 J_z \delta_{\bq,\bq'} +J_x  \bs{X}_{\bq} (\delta_{\bq,\bq'-\bs{n}_x} + \delta_{\bq-\bs{n}_x,\bq'}) \\ 
&+& J_y \bs{Y}_\bq (\delta_{\bq,\bq'-\bs{n}_y}+ \delta_{\bq-\bs{n}_y,\bq'})  \non \\
\Delta_{\bq \bq'} &=&  J_x \bs{X}_{\bq} ( \delta_{\bq,\bq'-\bs{n}_x} -\delta_{\bq-\bs{n}_x,\bq'}) \non \\     
&+& J_y \bs{Y}_\bq (\delta_{\bq,\bq'-\bs{n}_y}- \delta_{\bq-\bs{n}_y,\bq'}). 
\label{eq:xidelta}
\eea
and the non-zero entries of $H_1$ are given in FIG.~\ref{fig:Kitaev_extended}.  To specify the particular vortex configuration one, as before, replaces the operators $X_\bq$ and $Y_\bq$ by their eigenvalues in that configuration. On a torus, the $\xi$ and $\Delta$ given in (\ref{eq:xidelta}) are modified to include the terms that connect both sides of the torus, i.e.~the terms that connect the sites $(0,q_y)$ to $(N_x -1,q_y)$ and $(q_x,0)$ to $(q_x,N_y-1)$. The values of $X_\bq$ and $Y_\bq$ in these terms depend on the arrangement of vortices and the quantum numbers $l_{0}^{(x)}$ and $l_{0}^{(y)}$ of the two independent homologically non-trivial loop symmetries $L_{0}^{(x)}$ and $L_{0}^{(y)}$ that run through the origin $(0,0)$. The Hamiltonian for any sector on a torus can be now be generated by observing the following dependencies (see also FIG.~\ref{fig:ExplainF}):
\begin{equation}
\left\{ 
\begin{array}{lr}
{X}_{q_x,q_y} = \prod_{q_y'=0}^{q_{y}-1} W_{q_x,q_y'}&
(q_y \ne 0 \mathrm{~and~} q_x \ne N_x-1) \\
{X}_{q_x,q_y} = 1&
(q_y=0 \mathrm{~and~} q_x \ne N_x-1)  \\
X_{q_x,q_y} = - l^{(x)}_0 \prod_{q_y'=0}^{q_y-1}  W_{q_x,q_y} &
(q_y \ne 0 \mathrm{~and~} q_x=N_x-1) \\
X_{q_y,q_x}=-l^{(x)}_0&
(q_y = 0 \mathrm{~and~} q_x = N_x-1)\\ 
\end{array}
\right.
\end{equation}
\begin{equation}
\left\{ 
\begin{array}{lr}
Y_{q_x,q_y}=1& (q_y \ne N_y-1)\\
Y_{q_x,q_y}=-l^{(y)}_{q_x}& (q_y = N_y-1)\\
\end{array}
\right.
\end{equation}
where $l^{(y)}_{q_x} = l^{(y)}_0 \prod_{q_y=0}^{N_{y-1}} \prod_{q_x'=0}^{q_{x-1}}  W_{q_x',q_y}$. These values for $X_\bq$ and $Y_\bq$ can  be used for the extended Hamiltonian $H_1$ shown pictorially in FIG. \ref{fig:Kitaev_extended}   {\em except} for $P(\bq)^{(5)}$ and $P(\bq)^{(6)}$ when $q_y = N_y-1$. For $P(\bq)^{(5)}$, we have  
\begin{equation}
\left\{ 
\begin{array}{lr}
X_{q_x,q_y+1} = -l_{q_{x+1}}^{(y)}  & ( q_x \ne N_x-1)\\
X_{q_x,q_y+1} = l_0^{(x)} l_0^{(y)}  & ( q_x = N_x-1),\\
\end{array}
\right.
\end{equation}
while for $P(\bq)^{(6)}$,
\begin{equation}
\left\{ 
\begin{array}{lr}
X_{q_x,q_y} =l_{q_x}^{(y)} \prod_{q_y'=0}^{q_{y}-1} W_{q_x,q_y'}  & ( q_x \ne N_x-1)\\
X_{q_x,q_y} =  l^{(x)}_0 l_0^{(y)} W_{q_x,q_y}  & ( q_x = N_x-1).\\
\end{array}
\right.
\end{equation}

The system is diagonalized by solving the Bogoliubov-De Gennes eigenvalue problem
\bea
\label{bdgtransf}
\left[ \begin{array}{cc} \xi & \Delta \\ \Delta^\dagger & -\xi^{T} \end{array} \right] =  \left[\begin{array}{cc} U & V^* \\ V & U^* \end{array}  \right]  \left[ \begin{array}{cc} E & \bs{0} \\ \bs{0} & -E \end{array}  \right] \left[ \begin{array}{cc} U & V^* \\ V & U^* \end{array}   \right]^\dagger,
\label{eq:BdG}
\eea
where the non-zero entries of the diagonal matrix $E_{nm} = E_n \delta_{nm} $ are the the quasi-particle excitation energies. The Bogoliubov-Valentin quasi-particle excitations are 
\bea
\label{eq:gamma}
&&\left[\begin{array}{cc}\gamma_1^\dagger,...,\gamma_M^\dagger,  & \gamma_1,...,\gamma_M \end{array}  \right] \\&& =  \left[\begin{array}{cc} c_1^\dagger,..., c_M^\dagger, & c_1,...,c_M \end{array}  \right]  \left[\begin{array}{cc} U & V^* \\ V & U^* \end{array}  \right]  .
\eea
which after inversion and substitution into (\ref{eq:Hg}) give
\be
H= \sum_{\bs{n}=1}^{M} E_n (\gamma^\dagger_n \gamma^{\phantom \dagger}_n - \frac{1}{2}) .
\label{eq:Hd}
\ee
A short review of the relevant theory of the eigenstates and eigenvalues of Hamiltonians obtained in this way, in particular the application of the Bloch-Messiah-Zumino theorem, is given in the appendix. Using the prescriptions described there we can calculate the eigenstates and energies in all vortex and homology sectors on tori containing a few thousand spins on a desktop computer. In the vortex free sector however, there is considerable advantage to be gained by working in the momentum representation. The Hamiltonian in this case can be written as
\be
H_ = \sum_{\bk_x,\bk_y}  E_{\bk}( \gamma^\dagger_\bk
\gamma_\bk^{\phantom \dagger} - \frac{1}{2}),
\ee
where the dispersion relation $E_\bk$ is as given for the plane in
(\ref{eq:Ek}). The allowed values of $k_\alpha$ in the various homology sectors on the torus are $\theta_\alpha + 2 \pi
\frac{n_\alpha}{N_\alpha}$ for integer $n_\alpha=0,1,...N_\alpha-1$, where the four topological sectors, $(l^{(x)}_0,l^{(y)}_0)=(\pm 1,\pm 1)$ have values of $\theta_{\alpha}$ given by $\theta_\alpha = (\frac{l_0^{(\alpha)}+1}{2})\frac{\pi}{N_\alpha} $.
While it is simple to see that this expression is valid for the fully
periodic sector with $(l^{(x)}_0,l^{(y)}_0)=(-1,-1)$, it is not so
obvious how the general expression can be arrived at using the values for
for $X_\bq$ and $Y_\bq$ given above.
One way to understand this jump is to
imagine that the torus in question is doubled in period along the
$\alpha$ direction and the pattern of $X$'s and or $Y$'s is repeated
on the new part of the lattice, but with a $-1$ phase. On this new
lattice it is possible to define a fundamental domain of size $N_x
\times N_y$, such that all the values of $X_\bq$ and $Y_\bq$ are $+1$.
The periodic boundary conditions on the doubled torus now correspond
to anti-periodic boundary conditions on the fundamental domain and we
see that we can use the same dispersion relation as before, only with
shifted momenta.

One would naively expect that the ground-states of the Hamiltonian be given by the four analogues of the planar ground-state (\ref{eq:BCS}) corresponding to the four ground-states of the toric code, and with energies $-\frac{1}{2}\sum_{\bk}E_{\bk}$. These energies are not exactly equal for the four homology sectors on the torus, because the allowed momenta are different, but it is not difficult to see that, at least for the dispersion relations we have given, they approach each other rapidly as the system size is increased. 

However, there are two situations where the true ground-state of a topological and vorticity sector on the torus may not be given by the BCS product. First of all, there is a connection between the number of vortices and the number of fermions on the torus. A configuration with an odd number of vortices of electric type can only exist if there are an odd number of broken dimers, i.e.~an odd number of fermions. BCS products like eq.~(\ref{eq:BCS}) have even fermion number parity and hence do not apply to vorticity sectors which have an odd number of electric (or in fact magnetic) vortices. Secondly, there are situations where the allowed momenta together with the values of $u_{\bk}$ and $v_{\bk}$  make the expression for the BCS product state vanish. 

A particularly important example of this occurs in the vortex free sector with $(l^{(x)}_0,l^{(y)}_0)=(-1,-1)$. In the Abelian phase of the model, we can just use the BCS ground-state (\ref{eq:BCS}) as expected.  However  in the B-phase, we see that $\Delta_{\pi,\pi}=0$ and $\xi_{\pi,\pi}/E_{\pi,\pi}=-1$ implying that $u_{\pi,\pi}=0$ and $v_{\pi,\pi}=i$.  This cause the BCS state (\ref{eq:BCS}) to vanish because on a torus $c^\dagger_{\pi,\pi} c^\dagger_{-\pi,-\pi}=(c^\dagger_{\pi,\pi})^2=0$. It is important to note that this effect is not dependent on the B-phase being gapless, one only requires that $\xi_{\pi,\pi}$ is negative and $\Delta_{\pi,\pi}=0$. 

The vanishing of the BCS state is somewhat similar to what happens in Read and Green's treatment of the spinless p-wave superconductor \cite{Read00,RGnote}, but here, we cannot propose the expression in Eq.~(\ref{eq:occ_odd}) as an alternative because it has the wrong fermion number parity for the zero vortex sector.  However, one sees that
\be
\ket{\psi_{\pi,\pi}} = \prod_{\bk \neq (\pi,\pi)} (u_\bk +v_\bk c^\dagger_\bk
c^\dagger_{-\bk} ) \ket{ \{ W_\bq \},l^{(x)}_0,l^{(y)}_0, \{\emptyset \}},
\label{eq:gsnew}
\ee
with $(l^{(x)}_0,l^{(y)}_0)=(-1,-1)$ and $Q_\bq=1$ for all $\bq$, is an eigenstate with even fermion number, no vortex excitations and energy $-\sum E_\bk/2+E_{\pi,\pi}$. As with the generic situation shown in the Appendix, all of the states $\ket{\psi_{\bk'}}=\gamma^{\phantom\dagger}_{\pi,\pi} \gamma^\dagger_{\bk'} \ket{\psi_{\pi,\pi}}$ are also vortex-free eigenstates with energy $-\sum_\bk E_\bk/2+E_{\bk'}$. The ground-state of the vortex free system in this topological sector is precisely the state in this family for which $E_{\bk'}$ is minimal. If the system is gapped, then $E_{\bk'}$ does not approach zero even in the thermodynamic limit, and the ground-state of the  $(l^{(x)}_0,l^{(y)}_0)=(-1,-1)$ is gapped away from the degenerate ground-states of the other three vortex free sectors.  Even more generally, we can say that if the B-phase is gapped and the conditions $\xi_{\pi,\pi}/E_{\pi,\pi}=-1$ and $\Delta_{\pi,\pi}=0$ are fulfilled, the ground-state on a torus is three-fold degenerate, as expected from TQFT.

\section{Summary and outlook}
We have described a spin fermionization procedure for Kitaev's honeycomb model and related spin models.  Using this method we derived exact expressions for the ground-states and their associated eigenvalues. The derived ground-states are closed form expressions that do not require additional spectral projection. These expression combine two powerful wavefunction descriptors: the BCS product and the stabilizer formalism. The solution clarifies the nature of the topological phases of the model and the role the BCS product plays in determining them. It is clear now for example that the $\ZZ_2 \times \ZZ_2$  Abelian phase is determined from the fermionic vacuum, with the BCS product only adjusting this state slightly. As the vacuum is fixed and Abelian, the transition to the non-Abelian phase is therefore driven exclusively by the BCS product. 

We also showed how to extend our fermionization procedure to handle general vortex configurations on a torus and we discussed how the additional constraints due to the interdependence of loop symmetries and fermions arise in the calculations.  We closely examined the ground-state for the fully periodic vortex free sector, building on a more general discussion for arbitrary configurations given in the Appendix, and explained why the blocking mechanisms in the system dictate that the non-Abelian ground-state is three-fold degenerate, confirming the prediction from TQFT. We intend to use our fermionization to study degeneracies in systems with multiple anyonic excitations in the future and hope to elucidate the relation between the Abelian and non-Abelian anyons in the model.

While we have explicitly derived the relationship between the exact solution and the toric code ground-state we have not yet explored the relationship between the exact solution and its perturbative approximation.  This now possible, at least in principle, as the BCS product and the Brillouin-Wigner perturbation expansion both start from the TC ground-state. At the very least, the comparison should reveal the precise denominators in the perturbative expansion and clarify how this expansion breaks down at the phase transition. More speculatively, it may also help to extend perturbative techniques beyond the phase transition. The ability to do so could provide a new perspective on the non-Abelian phase and would be extremely useful for other models which are not exactly solvable. 

The results obtained on the torus highlight the connections between the blocking mechanisms, the ground-state degeneracy and the Abelian to non-Abelian phase transition. A more general analysis along these lines, which is not confined to a particular model would almost certainly be beneficial.

\begin{acknowledgments}
The authors would like to thank Ahmet Bolukbasi for discussions and clarifications.  This work has been supported by Science Foundation Ireland through the President of Ireland Research Award 05/Y12/1680 and the Principal Investigator Award 08/IN.1/I1961.
\end{acknowledgments}

\appendix
\section{Ground-state construction}
\label{app:HFB}

Here we review some of the relevant Hartree-Fock-Bogoliubov theory and
discuss how to bring the ground-states of each vortex sector into a
canonical form.  We restrict ourselves, as in the main text, to
situations where we have an even number of sites $M$ on the lattice,
and therefore an even number of $\gamma^{\dagger}$ excitations. For
more details we refer the reader to ref. \onlinecite{Ring04} on which much of
the following is based.

The ground-state for a fermionic Hamiltonian (\ref{eq:Hd}) can usually
be written down as a Hartree-Fock-Bogoliubov (HFB) projection,
\be
\ket{\text{gs}} = \prod_n \gamma_n \ket{\text{vac}}
\label{eq:HFB}
\ee
where the energy of this ground-state is $-\sum_n E_n /2$. However,
sometimes the physical situation will demand that the ground-state has
odd fermion number-parity $p$. In the honeycomb model for example the
fermionic number parity is completely determined form the vortex
configuration. Specifically we see that, because creating an e-m
vortex pair excitation necessary means breaking a z-dimer, the
fermionic number parity must be equal to the e-number parity and the
m-number parity.

The procedure in this case is to redefine our choice of
$\gamma^\dagger$ and $\gamma$ such that the lowest energy odd parity
state can be found. In practice we swap $\gamma^\dagger_1
\leftrightarrow \gamma_1$ and then set $ \ket{\text{gs}}_{odd} =
\gamma_1 \ket{\text{gs}}$.  This new state, annihilated by the
annihilation operators, has odd fermion number parity but has energy
$-\sum_n E_n /2 + E_1$ (Here our convention is to choose $E_1$ to be the smallest of the $E_i$).

While the above prescription handles a great many physical situations
there are a number of reasons why it is not always sufficient. The
first is simply that the method by which we construct the ground-state
is projective and therefore in many cases the vacuum state is not
uniquely defined. The second, and perhaps more important, is that
physical situations do exist such that the calculated $\gamma$
fermions are such that $\ket{\text{gs}}$ and therefore
$\ket{\text{gs}}_{odd}$ are already zero. A solution to both problems
can be found by making use of the Bloch-Messiah-Zumino theorem
\cite{Bloch62,Zum62}.

In practice the theorem says that we can do a singular value
decomposition of the  $M \times M$ matrix $U$
\be
U = D \mathcal{U} C
\ee
for unitary $C$ and $D$ such that the eigenvector matrix defining the
$\gamma^\dagger$ quasi-particle excitations, (\ref{eq:gamma}), can be
decomposed into
\be
\left[\begin{array}{cc} U & V^* \\ V & U^* \end{array}  \right]  =
\left[\begin{array}{cc} D & \bs{0} \\ \bs{0} & D^* \end{array}
\right]   \left[\begin{array}{cc} \mathcal{U} & \mathcal{V} \\
\mathcal{V} & \mathcal{U} \end{array}  \right]
\left[\begin{array}{cc} C & \bs{0} \\ \bs{0} & C^* \end{array}
\right]
\ee
where $\mathcal{U}$ and $\mathcal{V}$ take block diagonal forms

\bea
\mathcal{U} &=& \left[ \begin{array}{lllll}  Z & & & &   \\ &
\mathcal{U}_1 & & &   \\ & & \ddots & &     \\  & & & \mathcal{U}_n &
 \\  & & & &   I   \end{array} \right],
\label{eq:UBmat}
\eea
and
\bea
\mathcal{V} &=& \left[ \begin{array}{lllll}  I & & & &   \\ &
\mathcal{V}_1 & & &   \\ & & \ddots & &     \\  & & & \mathcal{V}_n &
 \\  & & & &   Z   \end{array} \right],
\label{eq:VBmat}
\eea
where $Z$ and $I$ are square zero and identity matrices respectively and
\bea
\mathcal{U}_i &=& \left[ \begin{array}{ll}  u_i & 0 \\ 0&u_i
\end{array} \right],
\label{eq:Umat}
\eea
and
\bea
\mathcal{V}_i &=& \left[ \begin{array}{ll}  0 & v_i \\ -v_i& 0
\end{array} \right].
\label{eq:Vmat}
\eea

\vspace{4mm}
This factorization means that (\ref{eq:gamma}) can be understood as
three separate transformations:
\vspace{3mm}

\nin (1) A unitary operator that mixes the fermionic excitation and
annihilation operators amongst themselves:
\be
a_i^\dagger = \sum_j D_{ji} c^\dagger_j \quad  a_i =\sum_j D_{ji}^* c_j.
\ee
This defines what is known as the canonical basis.
\vspace{3mm}

\nin (2)  A Bogoliubov Transform which, for paired levels $(u_k >0 ,
v_k >0)$, mixes creation and annihilation operators
\be
\alpha_k = u_k a^\dagger_k -v_k a_{\bar{k}}  \quad \alpha_{\bar{k}} =
u_k a^\dagger_{\bar{k}} +v_k a_{k}
\ee
where the $(k,\bar{k})$ label elements of the $2 \times 2$ matrices
(\ref{eq:Umat}) and (\ref{eq:Vmat}), and for blocked levels which are
either occupied $(v_i=1, u_i=0)$ or empty $(v_j=0, u_j=1)$:
\bea
\non
\alpha_i^\dagger &=& a_i ,\;\; \alpha_j^\dagger = a_j^\dagger, \\
\alpha_i &=& a_i^\dagger ,\;\; \alpha_j = a_j .
\eea

\vspace{3mm}
\nin (3) A unitary operator which mixes quasi-particle operators
$\alpha^\dagger_k$ amongst themselves:
\be
\gamma^\dagger_l = \sum_{k} C_{kl} \alpha^\dagger_{k} \quad \gamma_l =
\sum_{k} C_{kl}^* \alpha_{k}.
\ee

\vspace{4mm}

The ground-state, written in the canonical basis, is
\be
\ket{\text{gs}} = \prod_i^m a_i^{\dagger} \prod_{k \ne i} (u_k +v_k
a^\dagger_k a^\dagger_{\bar{k}}) \ket{\text{vac}},
\label{eq:occ_odd}
\ee
where the first product is over the $m$ occupied levels only and it is
understood that all $u$'s in the second product are non-zero.
The state is annihilated by all $\gamma_n$ and thus has an energy
$-\sum_n E_n/2$. To see this first note that the state is annihilated
by all $\alpha_{i}$ and then that each $\gamma_l$ is a linear superposition
of these.

The fermion number-parity $p$ of the state is dictated by the number
of occupied modes $m$. The appearance of an odd number of occupied
modes implies that the final state is of different number parity to
the vacuum. This is exactly the situation observed by Read and Green
for the spinless p-wave superconductor on a torus
\cite{Read00,RGnote}. In the honeycomb lattice model on a torus, the
situation is complicated slightly by the fact that the fermion number parity of an eigenstate is determined exclusively by the vortex configuration sector to which it belongs. This immediately implies that an odd (even) number of occupied modes is not
allowed in a vortex sector with even (odd) e-number or m-number parity. When
these situations occur we say that the state (\ref{eq:occ_odd}) is
blocked and we must re-arrange the eigenvector matrix, effectively
switching the $\gamma^{\dagger}_{i}$ and $\gamma^{\dagger}_{i}$, such that the eigenstate with the lowest energy is achieved.  Below we discuss t he four possible scenarios
and describe how to construct the eigenstate in each case.

\vspace{4mm}

\nin(A) Even $p$, Even $m$:

\vspace{2mm}

Here we may use (\ref{eq:occ_odd}) with no modification. The
ground-state energy is $E_{\text{min}}=-\sum_n E_n/2$. One encounters
this situation in the vortex free sector in both the A-phase and the
three partially/fully anti-periodic sectors of the B-phase.

\vspace{2mm}

\nin(B) Even $p$, Odd $m$:
\vspace{2mm}

Here we also use (\ref{eq:occ_odd}) but the singular value
decomposition to calculate the $u$'s and $v$'s is performed after
first switching columns $(U_{l1},V_{l1}) \leftrightarrow
(V_{l1}^*,U_{l1}^*)$. Of course an initial singular value
decomposition of the original $U$ matrix is first needed in order to
determine the number of occupied modes $m$. The switching of columns
of the matrix effectively changes an occupied mode for an empty one
and the energy of the ground-state is therefore $E_{\text{min}}=-
\sum_n E_n/2 +E_1$.  We see that if the sector has even fermion parity
and is gapped, an odd number of occupied modes $m$ implies a raising
of the energy above what one might otherwise expect. We encounter this
situation in the B-phase of the vortex-free sector with
$(l^{(x)}_0,l^{(y)}_0)=(-1,-1)$.
\vspace{2mm}

\nin(C) Odd $p$, Even $m$:
\vspace{2mm}

This situation is again handled by switching occupied and empty modes
$(U_{l1},V_{l1}) \leftrightarrow (V_{l1}^*,U_{l1}^*)$ and the energy
of the ground-state is again $E_{\text{min}}=-\sum_n E_n/2 +E_1$. The
vacuum in this case must be from the vortex sector such that the
operation of the {\em now} odd number of $a^\dagger_i$'s gives the
vortex and topological sector for which we calculated the $U$ and $V$
matrices.

\vspace{2mm}
\nin(D) Odd $p$, Odd $m$.
\vspace{2mm}

The ground-state in this sector is given by (\ref{eq:occ_odd}) without
modification and the ground-state energy is $-\sum_n E_n/2$. Thus, if
the sector is gapped, the energy of this ground-state is lower than
what one might expect from an odd fermion number state.  As in the
previous case the vacuum must be defined so that operating with an odd
number of $a^\dagger_i$'s gives the correct vortex and topological
sector.

\vspace{4mm}
In all of the above situations, because the $\gamma_i^\dagger$'s do
not commute with all loop symmetries,  the excited states in each
sector are obtained by operating on the ground-state with quadratic
operators $\gamma^{\dagger}_i \gamma^\dagger_j$. This can also be
checked through a simple counting argument. On a torus, as we have
$M+1$ independent loop symmetries, the Hilbert space dimension of each
sector is $2^{M-1}$, see \cite{Kel08a}. If we were allowed to operate
with single $\gamma^\dagger_i$'s and $\gamma_i$'s on the ground-state
we could generate $2^M$ states, which is obviously too many.


\begin{thebibliography}{99}
\bibitem{RGnote} In \cite{Read00} it was noted that in the weak pairing phase
the term $u_{\bs{0}} + v_{\bs{0}} c^\dagger_{\bs{0}} c^\dagger_{\bs{0}}= 0$ and
thus (\ref{eq:BCS}) vanishes on a torus. The ground-state for this sector is $\ket{\text{gs}} = \prod_{\bk \neq \bs{0}} (u_\bk + v_\bk \crc_\bk \crc_{-\bk}) c^{\dagger}_{\bs{0}} \ket{ \text{vac} }$.
\end{thebibliography}
\end{document}